\documentclass{emulateapj}

\usepackage{subfigure}
\usepackage{times}
\usepackage{graphicx}
\usepackage{verbatim} 
\usepackage{tabularx}

\newcommand\T{\rule{0pt}{2.6ex}}
\newcommand\B{\rule[-1.2ex]{0pt}{0pt}}
\newcolumntype{Z}{>{\centering\arraybackslash}X}

\shorttitle{\emph{Chandra} observations of PSR J1809$-$2332}
\shortauthors{Van Etten et al.}

\begin{document}

\title{A \emph{Chandra} Proper Motion for PSR J1809$-$2332}

\author{Adam Van Etten, Roger W. Romani}
\affil{Department of Physics, Stanford University, Stanford, CA 94305}
\email{rwr@astro.stanford.edu}
\author{ \& C.-Y. Ng\altaffilmark{1}}
\affil{Department of Physics, McGill University, Montreal, QC H3A 2T8, Canada}
\altaffiltext{1}{Tomlinson Postdoctoral Fellow, CRAQ Postdoctoral Fellow}


\begin{abstract}

We report on a new \emph{Chandra} exposure of PSR J1809$-$2332, the recently 
discovered pulsar powering the bright EGRET source 3EG J1809$-$2328. By
registration of field X-ray sources in an archival exposure, we measure
a significant proper motion for the pulsar point source over an $\approx 11$ year baseline.
The shift of $0.30 \pm 0.06\arcsec$ (at $PA= 153.3\pm18.4$) supports an
association with proposed SNR parent G7.5$-$1.7. Spectral analysis of
diffuse emission in the region also supports the interpretation as a hard 
wind nebula trail pointing back toward the SNR.
\end{abstract}

\section{Introduction}

3EG J1809$-$2328 was one of the brightest unidentified hard-spectrum gamma-ray 
sources detected by EGRET \citep{hartman99}. ASCA observations of the EGRET error box revealed
an extended X-ray source \citep{rrk01}, suggestive of a pulsar wind nebula (PWN).
A 9.7\,ks \emph{Chandra} ACIS exposure in August 2000 \citep{braje02} revealed 
a point source connected to the non-thermal X-ray/radio nebula, bolstering the PWN identification.  
\citet{roberts08} then described G7.5$-$1.7, a partial $\sim0.5-0.8^{\circ}$ radius radio 
shell and possible supernova remnant; the center of this SNR candidate lies $0.3^{\circ}$
from the point source, in the general direction defined by the PWN trail.

The PSR/PWN nature of the source was confirmed by the discovery of PSR J1809$-$2332,
a $P=147$\,ms, $\tau_c=68$\,ky pulsar discovered by blind search of the 
\emph{Fermi} Large Area Telescope (LAT) $\gamma$-ray photons \citep{abdo09}.
This energetic (${\dot E} = 4.3 \times 10^{35} {\rm erg\,s^{-1}}$) pulsar powers
the $\gamma$-ray and PWN emission and has a timing position consistent with the X-ray
point source at the PWN apex \citep{ray11}. As for most $\gamma$-ray selected pulsars, we
lack a dispersion measure estimate of the distance. However, \citet{oka99} suggested
that the diffuse X-ray emission is anti-correlated with molecular gas in the Lynds 227 
dark cloud; if associated this implies a plausible $d\sim$ 1.8 kpc.

	To test the SNR association and to improve spectral and morphological measurements
of the diffuse emission, we have obtained a new \emph{Chandra} ACIS exposure. By
carefully matching to the original pointing, we have minimized systematic effects
and allowed excellent frame-referencing and astrometry of the point source. The 
$\sim 11$\,y baseline between the exposures thus allows a sensitive study of the 
proper motion and origin of PSR J1809$-$2332.

\begin{figure*}[ht!]
\epsscale{1.1}\label{fig:f1}
\centering
\subfigure[]{
\includegraphics[width=0.49\textwidth]{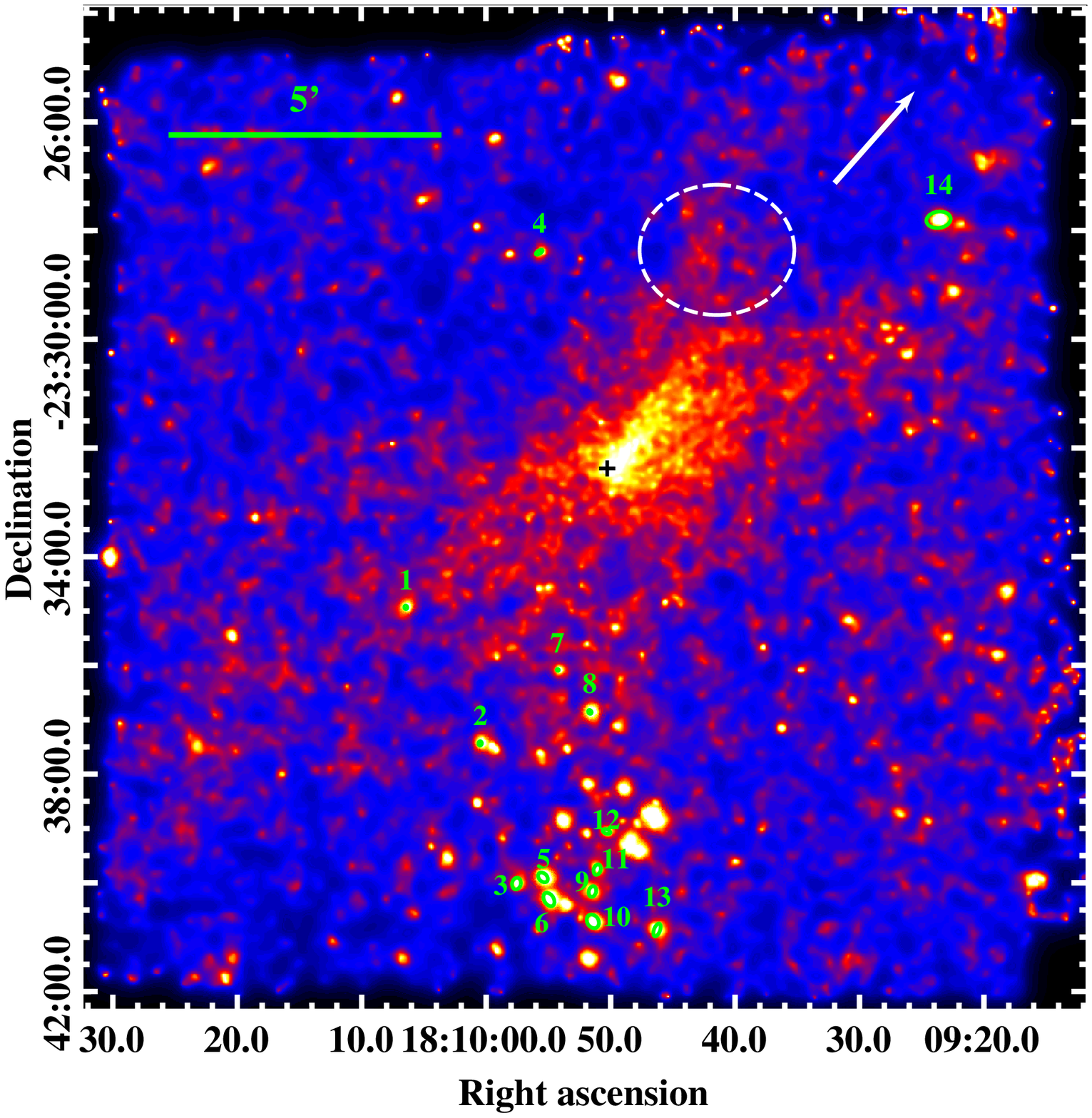}
\label{fig:f1a}
}
\centering
\subfigure[]{
\includegraphics[width=0.49\textwidth]{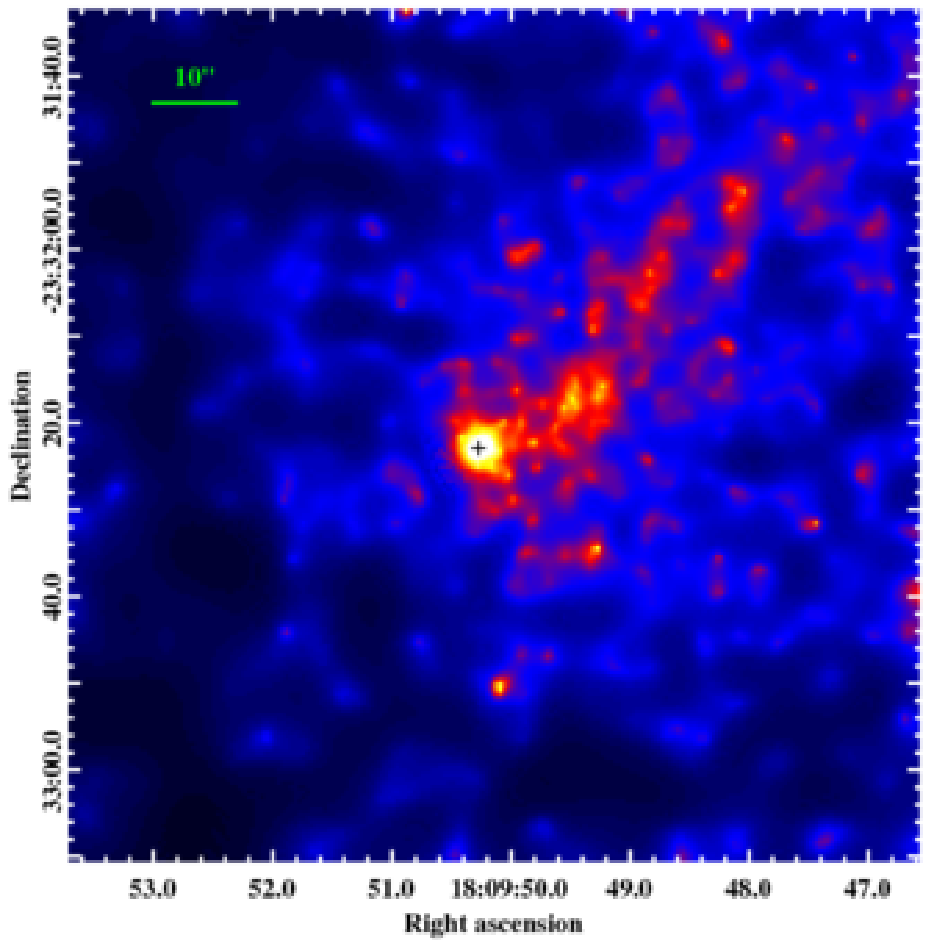}
\label{fig:f1b}
}
\caption{
Left: Merged \emph{Chandra} exposure corrected $0.5-7$ keV ACIS-I array image 
of the extended emission around PSR J1809$-$2332 (observation ID 739, 12546), 
adaptively smoothed (0.5 to 15 pixel Gaussian kernels, 
$10^{-6} {\rm erg\, cm^{-2}s^{-1}}$ flux minimum).  
A possible bowshock and trail are seen extending northwest from the pulsar.
Point sources used in frame matching are shown in green, with the pulsar position
marked with a black cross. The white ellipse denotes the $1\sigma$ error 
ellipse for the extrapolated pulsar position some 10 kyr in the past, while
the white arrow points to the proposed explosion site for SNR G7.5$-$1.7.  
These two directions agree within $1\sigma$.
Right: Merged adaptively smoothed (0.5 to 15 pixel Gaussian kernels, 3 count minimum)
0.5-7 keV image of the $100\arcsec$ surrounding the pulsar. Both panels are displayed
with a logarithmic stretch.
}
\end{figure*}

\section{X-Ray Data Analysis}

PSR J1809$-$2332 was observed on July 28, 2011 (obsID 12546) for 29.7 ks in
timed VFAINT mode. As for our August 16, 2000 exposure (obsID 739) we used
the ACIS-I array. To minimize changes in the reference star PSFs (see below)
we matched the roll angle to the 2000 epoch. The point source position was not,
of course, known in the initial exposure and fell rather close to a
chip boundary. We thus shifted the pointing $\sim 30^{\prime\prime}$ in each coordinate
to move the point source further onto the I3 chip and improve measurement of diffuse
emission in its vicinity.
The data were analyzed with CIAO 4.3, after re-processing both frames to the most
current ACIS calibration.  No flaring occurred during obsID 12546, so the 
full 29.7\,ks was used. For obsID 739, the full 9.7\,ks was also usable.

	After combining, exposure correcting (with {\it fluximage}) and 
adaptive smoothing, the ACIS-I
frame gives a good overall view of the pulsar vicinity (Figure 1). There is
appreciable diffuse X-ray emission across the center of the image. The pulsar
point source lies near the apex of a patch of brighter nebulosity that trails off
to the North-West for $\sim 5^\prime$. On smaller scales the PWN has a
bright core behind the pulsar, $\sim 3-5\arcsec$ extension transverse to this trail
at the point source and fainter diffuse emission `ahead' of the pulsar. To the 
South of the pulsar, numerous bright X-ray sources from the young star cluster
S32 are seen. These bright stars provide excellent sources to aid frame registration
between epochs. There are some low significance X-ray concentrations outside of the bright
inner nebula in Figure \ref{fig:f1b}. None have obvious stellar counterparts
so it is unclear if these are background sources or clumps in the diffuse nebular
emission. 

\begin{table*}[!!ht]
\caption{Wavdetect Source Positions}
\label{tab:sources}
\centering
\begin{tabularx}{1.0\textwidth}{lZZZZZZZ}
\hline \hline
Src \T \B & Obs1 X & Obs1 Y & Obs1 Sig & Obs2 X & Obs2 Y & Obs2 Sig\\
\hline
1\T   & $3727.7\pm0.2$ & $3830.9\pm0.3$ & 8.9   & $3796.6\pm0.1$ & $3892.5\pm0.1$ & 34.3\\
2   & $3894.4\pm0.2$ & $3527.3\pm0.3$ & 35.4  & $3963.4\pm0.3$ & $3587.3\pm0.4$ & 6.0\\
3   & $3971.2\pm0.7$ & $3213.6\pm0.8$ & 5.9   & $4044.7\pm0.5$ & $3272.7\pm0.6$ & 11.6\\
4  & $4026.8\pm0.4$ & $4628.9\pm0.5$ & 5.4   & $4095.9\pm0.4$ & $4689.4\pm0.3$ & 11.9\\
5   & $4035.3\pm0.4$ & $3225.6\pm0.4$ & 24.4  & $4104.2\pm0.2$ & $3286.5\pm0.2$ & 42.4\\
6   & $4049.1\pm0.5$ & $3177.4\pm0.4$ & 24.1  & $4116.4\pm0.4$ & $3238.1\pm0.5$ & 15.7\\
7   & $4068.0\pm0.2$ & $3691.5\pm0.3$ & 6.1   & $4137.6\pm0.2$ & $3752.1\pm0.2$ & 11.8\\
8   & $4140.3\pm0.2$ & $3597.1\pm0.2$ & 31.7  & $4209.4\pm0.1$ & $3657.8\pm0.1$ & 51.0\\
9  & $4144.9\pm0.9$ & $3194.2\pm0.9$ & 6.0   & $4215.1\pm0.5$ & $3254.8\pm0.8$ & 6.0\\
10   & $4149.2\pm0.5$ & $3125.5\pm0.5$ & 17.8  & $4216.1\pm0.7$ & $3187.9\pm0.8$ & 8.2\\
11  & $4153.8\pm1.0$ & $3244.0\pm0.9$ & 5.7   & $4225.9\pm0.6$ & $3303.8\pm0.8$ & 6.1\\
12  & $4176.4\pm1.1$ & $3327.8\pm1.0$ & 5.2   & $4246.8\pm0.6$ & $3389.5\pm0.4$ & 8.5\\
13   & $4289.4\pm1.2$ & $3108.4\pm1.0$ & 7.3   & $4360.0\pm0.5$ & $3168.4\pm0.9$ & 5.2\\
14  & $4922.8\pm0.9$ & $4702.8\pm1.1$ & 7.6   & $4991.8\pm0.7$ & $4762.0\pm0.5$ & 17.7\\
psr & $4179.65\pm0.12$&$4143.61\pm0.13$&29.5  & $4248.44\pm0.06$&$4204.07\pm0.06$ & 59.1\\
\hline
\end{tabularx}
\end{table*}

\section{Frame Pointing Offset}

	To best constrain any pulsar proper motion, we must optimally register
the two observation epochs, using field point sources. We identify these using the CIAO 
wavdetect tool, selecting sources with a significance $\ge$5 in both observations.  
The young S32 star cluster $\sim8\arcmin$ due South of the pulsar provides
the bulk of the reference stars; inevitably several bright stars fall 
into chip gaps in one or the other of the observations, precluding their use.
This leaves 14 stellar sources, plus the pulsar.  We label the reference stars by increasing physical 
pixel x-coordinate (decreasing RA), mark these in  Figure \ref{fig:f1a} and tabulate the
wavdetect pixel positions and significances for each epoch in Table \ref{tab:sources}.

The wavdetect tool uses "Mexican hat" wavelet functions of different scales to
generate a source list and estimate position. This approximation may be inadequate
far from the ACIS aimpoint where the mirror PSF induces substantial and systematic
changes in the point source count distribution. By approximately matching the pointing
and roll angle we have kept the change in such distortions minimal between the two frames.
Nevertheless the PSF shape changes across the frame are very large and 
the majority of our reference stars lie $\ge 6^\prime$ from the aimpoint and pulsar.
We have improved the astrometric accuracy and further reduced the effect of PSF variation
by creating a simulated PSF for each point source and using these to fit the source position.

\begin{figure*}[htb!]
\label{fig:f2}
\centering
\subfigure[]{
\includegraphics[width=0.3\textwidth]{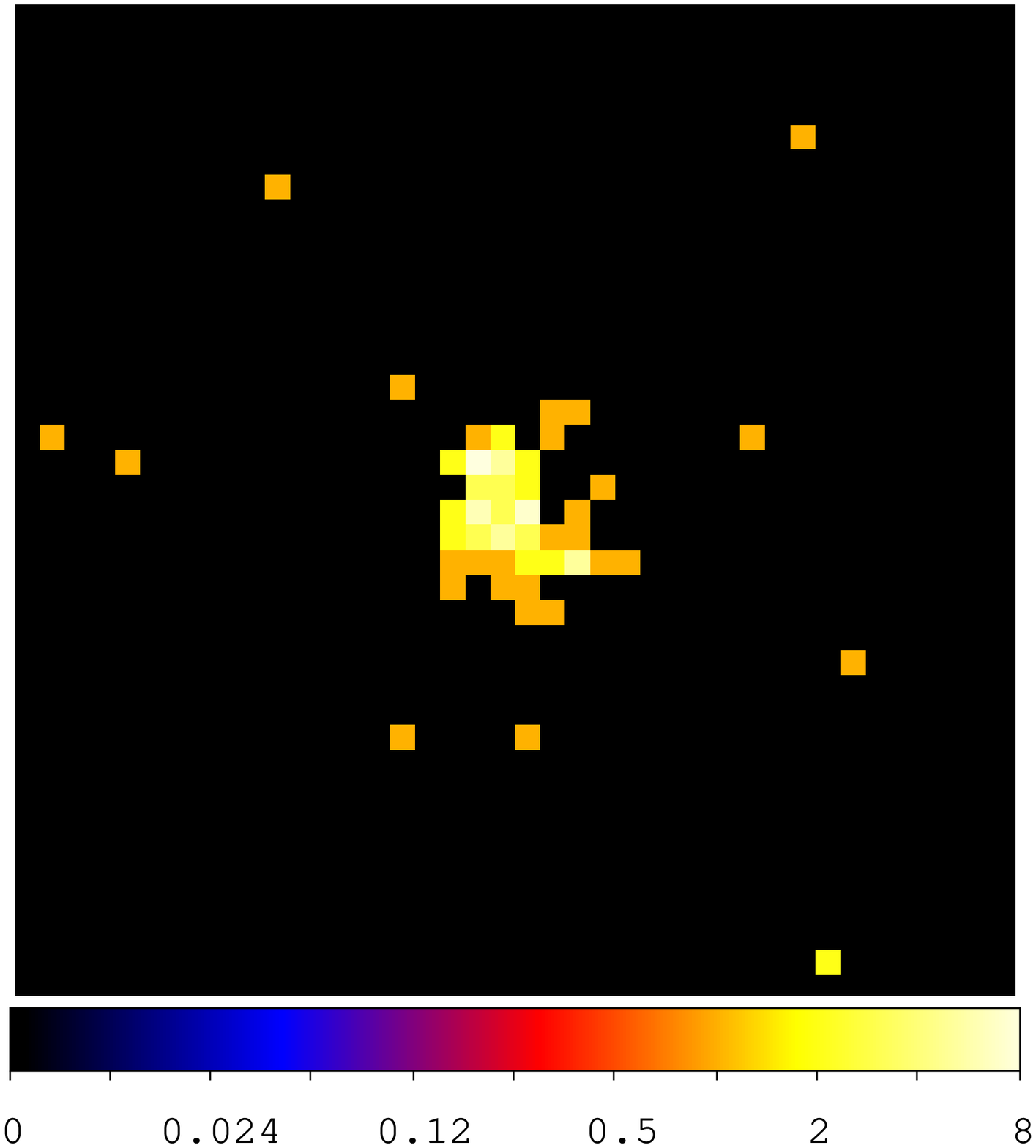}
\label{fig:src7evt}
}
\centering
\subfigure[]{
\includegraphics[width=0.3\textwidth]{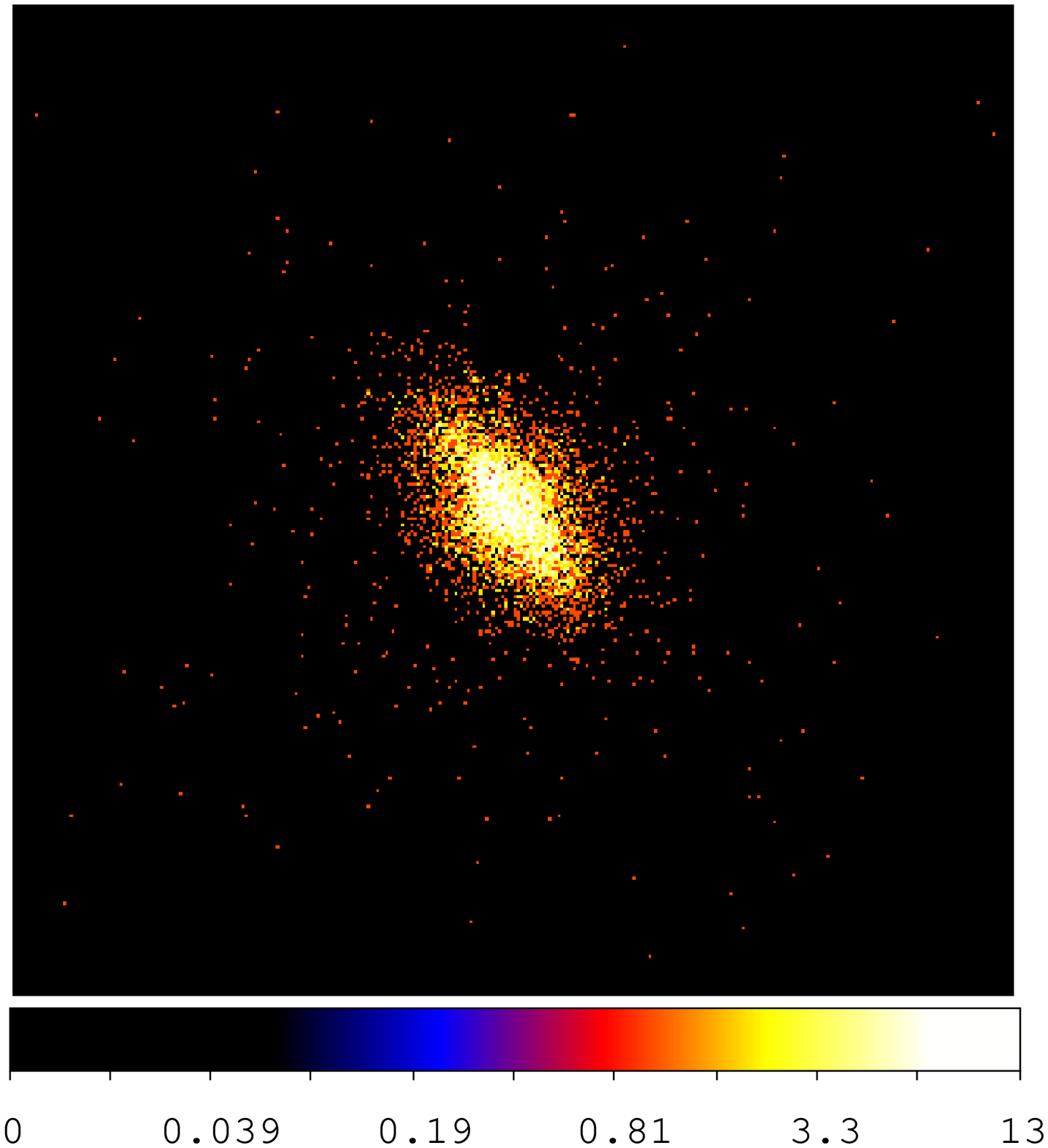}
\label{fig:src7psf}
}
\centering
\subfigure[]{
\includegraphics[width=0.3\textwidth]{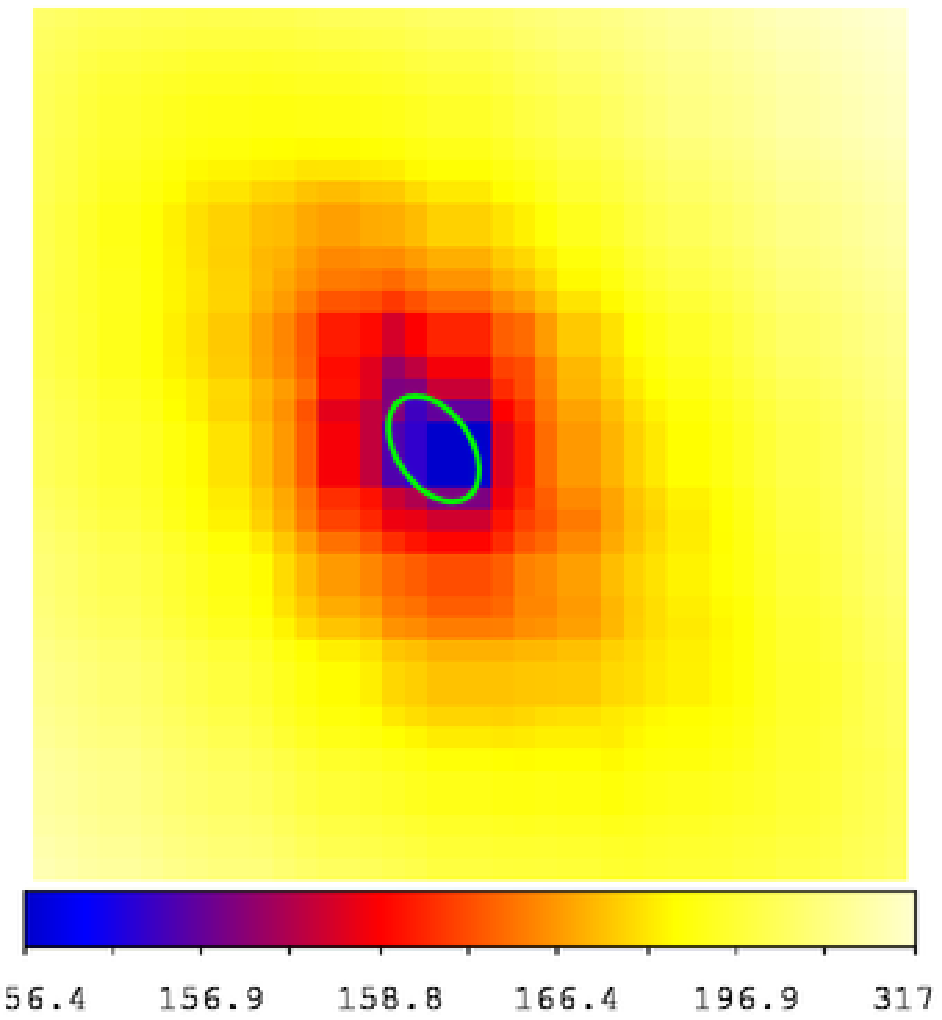}
\label{fig:src7fom}
}
\caption{
(a) \emph{Chandra} events cutout of Source 8 (2011 epoch).
(b) Simulated PSF at this source position.
(c) Computed FoM map zoomed 10$\times$, the green $\delta$FoM = 0.5 contour is
shown. The extrema in $x$ and $y$ of this ellipse give the positional uncertainty
in the source position.
}
\end{figure*}

We extract the spectrum of each source with the specextract function, fitting the stellar 
point sources to an absorbed Mekal plasma model, and save this spectrum within the Sherpa 
plotting program with the save$-$chart$-$spectrum command. After re-scaling
by $100\times$ to improve PSF modeling statistics, the spectra are passed to 
CHaRT\footnotetext{http://cxc.harvard.edu/chart/runchart.html},
which is a web interface to the SAOsac raytrace code.  CHaRT also takes as input
the exposure time as well as the source position, which is assumed to be the 
wavdetect position.  The output of CHaRT is subsequently input into MARX to create
model PSF events files at each source position, corrected for the offset of
the science instrument module (SIM) from the nominal location.   

	The modeled source events are binned to $1/8$ of the native ACIS pixel
resolution and recorded in 320$\times$320 image files. These are compared with
data cut-outs with matched 40$\times$40 native-pixel images. PSF models and data
cutouts are prepared for each star in both frames. We use these to solve for PSF
{\it vs.} data position shifts. The fitting program uses a maximum likelihood `figure
of merit' (FoM, the negative ln Poisson probability of getting the observed data
counts from the model PSF), where we run a grid of trial positions shifted
at the 1/8-pixel PSF model resolutions and sum the model counts to the observed ACIS
pixel. The result is a map of source likelihood in PSF position shift $x$ and $y$
(relative to the nominal wavdetect location). An example of the cut-out, PSF model
and shift map are shown in Figure 2.

In order to compute the best-fit position of the source, we fit an elliptic
paraboloid to the FoM surface.  The minimum of this surface determines the best
fit offset, and the offset error is estimated from the $\delta$FoM contours.
As with Source 8, several of the ellipses had large axis ratios. However an attempt 
to improve registration by fitting along major and minor axis did not reduce the 
scatter. Thus to be conservative, we project the ellipses to the $x$ and $y$ 
pixel axes and fit in this unrotated space.
We determine that $1\sigma$ (wavdetect) errors correspond to a $\delta\rm{FoM} = 0.5$, 
and thus ascribe $1\sigma$ error ellipses to the region enclosed by this
FoM increase above the fit minimum. Comparison with sources near the aimpoint,
where a circular Mexican hat should be an adequate approximation, confirm that
this provides a good error estimate. A typical fit and uncertainty are shown
by the green ellipse in Figure 2.

Table 2 gives the PSF-fit computed pixel offset of each source from the 
wavdetect position, along with estimates for the position errors.
The mean offset of all reference sources ($\approx 0.05$ pixels) is not significantly different 
from zero in either axis, showing that the wavdetect solutions do not impose
any large systematic offset between our matched exposures.

\begin{table}[h!]
\label{tab:fomsh}
\caption{PSF Fit Offsets from Wavdetect Positions}
\begin{tabular}{lcc}
\hline
\hline 
Src \T \B & Obs1 (X, Y) & Obs2 (X, Y)\\
\hline
1 \T& $(     -0.17 \pm       0.26, \;       +0.53 \pm       0.38)$ & $(      +0.08 \pm       0.12, \;      -0.18 \pm       0.15)$ \\ 
2 & $(     -0.01 \pm       0.18, \;      -0.16 \pm       0.24)$ & $(     -0.22 \pm       0.52, \;       +0.28 \pm       0.67)$ \\ 
3 & $(      +1.61 \pm       1.68, \;      -1.59 \pm       1.46)$ & $(     -0.09 \pm       0.44, \;      -2.05 \pm       0.85)$ \\ 
4 & $(      +0.22 \pm       0.70, \;       +0.19 \pm       0.62)$ & $(      +0.30 \pm       0.46, \;       +0.35 \pm       0.37)$ \\ 
5 & $(      +0.03 \pm       0.24, \;      -0.26 \pm       0.33)$ & $(      +0.38 \pm       0.18, \;      -0.09 \pm       0.22)$ \\ 
6 & $(     -0.61 \pm       0.37, \;      -0.24 \pm       0.39)$ & $(      +0.62 \pm       0.51, \;      -0.24 \pm       0.49)$ \\ 
7 & $(      +0.50 \pm       0.35, \;      -0.77 \pm       0.44)$ & $(     -0.34 \pm       0.28, \;      -0.13 \pm       0.32)$ \\ 
8 & $(     -0.09 \pm       0.20, \;      -0.06 \pm       0.23)$ & $(      +0.05 \pm       0.13, \;       +0.11 \pm       0.14)$ \\ 
9 & $(      +1.36 \pm       0.24, \;       +0.07 \pm       0.36)$ & $(     -0.39 \pm       0.29, \;       +0.17 \pm       0.33)$ \\ 
10 & $(     -0.55 \pm       0.36, \;       +0.10 \pm       0.46)$ & $(      +1.83 \pm       0.85, \;      -1.46 \pm       0.83)$ \\ 
11 & $(      +0.65 \pm       0.33, \;      -0.51 \pm       0.54)$ & $(     -0.96 \pm       0.32, \;       +1.01 \pm       0.28)$ \\ 
12 & $(     -0.38 \pm       0.73, \;       +0.68 \pm       0.82)$ & $(     -0.96 \pm       0.54, \;      -0.46 \pm       0.45)$ \\ 
13 & $(     -1.20 \pm       0.92, \;       +1.40 \pm       0.96)$ & $(     -0.98 \pm       0.64, \;       +1.24 \pm       0.60)$ \\ 
14 & $(      +1.16 \pm       0.37, \;      -1.05 \pm       0.39)$ & $(      +0.98 \pm       0.27, \;       +0.02 \pm       0.21)$ \\ 
psr \B & $(    -0.16 \pm       0.12, \;      -0.11 \pm	  0.12)$ & $(      +0.04 \pm       0.06, \;      -0.15 \pm       0.06)$ \\

\hline
\end{tabular}
\end{table}

We can now use our stellar position estimates to register the frames.
For the wavdetect positions we use the values in Table \ref{tab:sources}
directly. For PSF-fits we add the additional offsets of Table 2
and use PSF localization errors. For each source $i$, we define, e.g., $x_i$ as the
difference in $x$ coordinate between old and new exposure; epoch positional errors 
are added in quadrature to give $\delta x_i$.  The best fit frame shift is 
determined by minimizing the $\chi^2 = \sum_{i=1}^{14}[(x_i - S_x)/\delta x_i]^2$ 
between frames, where $S_x$ is the frame shift.  Similar values are computed for
the y coordinate. 

Inspection of the tables shows that source 3 has both a
large shift between frames and a large PSF-fit error. We suspect that it may
be a confused double with variable components (hence the poor PSF fit and large
apparent shift). We thus omit this star from the registration, leaving
13 field stars. We also attempted recursive pruning of field stars with the largest 
remaining $\delta x_i/\sigma_{x_i}$ or $\delta y_i/\sigma_{y_i}$; as expected 
the nominal fit errors decreased slightly
(especially for the PSF-fit solution). However the shifts were small and
decreasing the number of fit stars may increase sensitivity to systematics.
We thus conservatively retain all stars except the obvious outlier source 3
and report the best-fit shifts and shift errors in Table \ref{tab:frameshift}.
Although many of our reference stars are from the S32 cluster, comparison 
with the non-cluster stars does not show any large systematic shifts.
This is as expected since typical $\sim 10$km/s association velocities 
are small compared to the expected pulsar space velocity, as is the velocity 
due to differential Galactic rotation, for the expected distances.

\begin{figure}[h!]
\epsscale{1.1}
\plotone{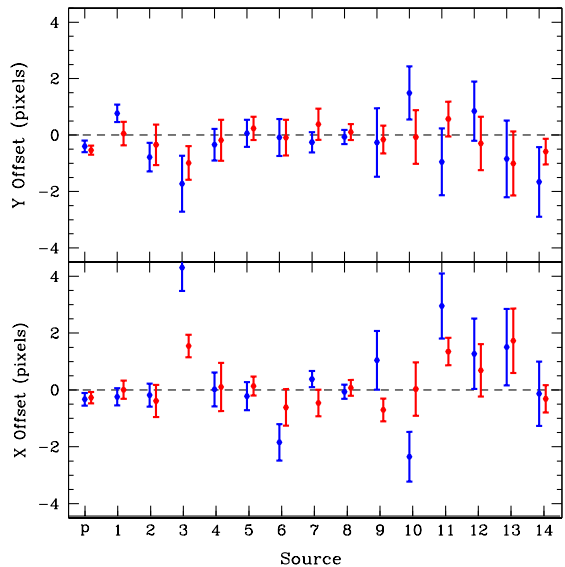} 
\label{fig:pos2}
\caption{
Residual source displacements between epochs ($\delta = $ new position $-$ old position), 
relative to the PSF-fit frame offset (Table \ref{tab:frameshift}).
Blue indicates wavdetect positions and errors, while red indicates
positions and errors obtained from the FoM maps. Source 3, an obvious outlier, is
pruned from the frame-shift fits. The pulsar shows small but significant shifts.
}
\end{figure}

Table \ref{tab:frameshift} lists the best-fit frame shifts for both
the raw wavdetect positions and the PSF fit positions. In both cases, we
compute shifts only, adopting the {\it CXO}-determined roll angle.
The net pixel shifts are consistent between the two methods. However, the
$\chi^2$ and frame offset errors are significantly smaller for the PSF-fit
approach. Figure \ref{fig:pos2} shows why: the scatter of the PSF-fit position
offsets about the best-fit line is smaller, with the exception of the obvious
outlier Source 3. We thus adopt the PSF-fit measurements and offset. These
finally allow us to compute the position shift (relative to the PSF-fit registration)
of the pulsar between the two frames. The values for both wavdetect and improved
PSF-fit detect centroiding are shown in Table \ref{tab:psroffset}.  This table 
also gives the best fit radial (in native ACIS pixel units) 
and PA shifts with propagated errors. 
In Figure 4 we show the registered frames near the pulsar,
after shifting the 2000 epoch by $(69.17, \, 60.83)$ pixels and regridding
with the CIAO dmregrid2 function.
The circle marks the best-fit 2000 position while the ellipse marks the
2011 epoch localization, including frame shift uncertainties.  Between epochs the pulsar
shows a significant $4.6\sigma$ shift. We thus have detected the proper motion of 
PSR J1809$-$2332; we comment on the implications below.

To determine the absolute position at the 2011 epoch, we computed a least-squares fit
using the PSF-fit positions and the optical positions of their USNO B-1 counterparts
to obtain a 2011.66 pulsar location of RA(2000.0)=18:09:50.25$\pm 0.03$, 
DEC(2000.0)=$-$23:32:22.68$\pm 0.10$.

\begin{table}[h!]
\label{tab:frameshift}
\caption{Frame Offsets: Two Fit Methods}
\centering
\begin{tabular}{lcccc}
\hline \hline 
Method \T \B & Coord. & Shift (pix) & $\chi^2/DoF$ \\
\hline
Wavdetect\T&X&	$69.12\pm0.18$&	29.9/12\\
Wavdetect & Y&	$60.86\pm0.15$&	16.2/12\\
PSF Fit &	X&	$69.17\pm0.14$&	18.4/12\\
PSF Fit\B&Y&	$60.83\pm0.09$&	5.0/12\\
\hline
\end{tabular}
\end{table}

\begin{table}[h!]
\caption{Pulsar Displacement (2011.66 -- 2000.71)}
\label{tab:psroffset}
\centering
\begin{tabular}{lccccc}
\hline \hline 
Method \T \B & X$^a$ & Y$^a$& R$^a$ (radial) & $\theta^b$  \\
\hline
Wavdetect\T&	$-0.33\pm0.23$&	$-0.40\pm0.21$ & $0.52\pm0.15$ & $140.6\pm24.2$\\
PSF Fit\B&	$-0.27\pm0.19$&	$-0.54\pm0.16$ & $0.60\pm0.13$ & $153.3\pm18.4$\\
\hline
\end{tabular}
$^a$ pixels (1 pixel = $0.49\arcsec$)

$^b$ measured in degrees CCW from north
\end{table}

\begin{figure}[h!]
\epsscale{1.05}
\plotone{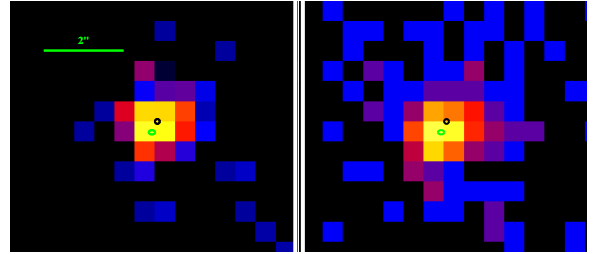}
\label{fig:psr_shift}
\caption{
Left: \emph{Chandra} image of the pulsar from the 2000 observation,  registered to the 2011 frame.
Right:  Matched \emph{Chandra} image from the 2011 observation.  Positions and errors (black circle
2000, green ellipse 2011) are taken from the FoM maps.  The 2011 uncertainty includes the frame-tie 
errors added in quadrature (Table \ref{tab:frameshift}).  The offset is significant at 
$\approx \! 4 \sigma$.
}
\end{figure}

\bigskip
\bigskip
\section{X-ray Spectra}

For our spectral study we define extraction regions for the several compact and extended
sources. On the smallest scales these are a $2\arcsec$ point source aperture
and a $3.5\arcsec \times 5\arcsec$ 'Torus' ellipse, transverse to the major nebula axis.
This small aperture is visible, but unlabeled
on Figure 5. On this figure we label the larger scale spectral extraction apertures:
a brightest nebula `Trail', a surrounding `Inner Nebula' and three
large regions of low surface brightness (an `Upstream' region ahead of the pulsar
and Northern and Southern extensions of the outer nebula). In each case the apertures
exclude any enclosed smaller scale region. As noted in Section 3 we also extracted
point-source spectra and fit absorbed Mekal models to the obvious stellar sources
to derive spectra for computing model PSFs. These stellar sources were not exceptional
and will not be discussed further here.  Background regions are defined 
on source-free portions of the I0, I1 and I3 chips; scaled backgrounds are subtracted from
the source spectra.

    Spectra are extracted with the CIAO version 4.3 specextract function, which also 
computes response files.  We group the spectra to a signal-to-noise/bin of 3, 
and fit all diffuse regions jointly with old and new-epoch data to an absorbed 
power-law model, using \emph{Sherpa}.  For the point source (pulsar) there is no 
evidence for variability, with $7.4 \pm 0.9 \times 10^{-3}$\,cps during the first epoch and  
$7.6 \pm 0.5 \times 10^{-3}$\,cps in the second. We therefore fit jointly to an
absorbed power-law plus a thermal component. Table \ref{tab:spec} gives the fit values and 68\% 
confidence errors. If we assume a common origin for the various components we can
improve the spectral constraints by fitting for a global $N_H$. Using the brightest
diffuse regions (Trail + Inner Nebula) we obtain 
$N_H = 1.3 \pm 0.2 \times 10^{22} {\rm cm^{-2}}$ 
and fix this for the other spectral regions. The Galactic HI surveys (nhtool)
show a column of $\sim 5 \times 10^{21} {\rm cm^{-2}}$ to $d \sim 0.5$\,kpc suggesting
a source distance $\ge1$\, kpc.

\begin{figure}[h!]
\label{fig:specregs}
\epsscale{1.2}
\plotone{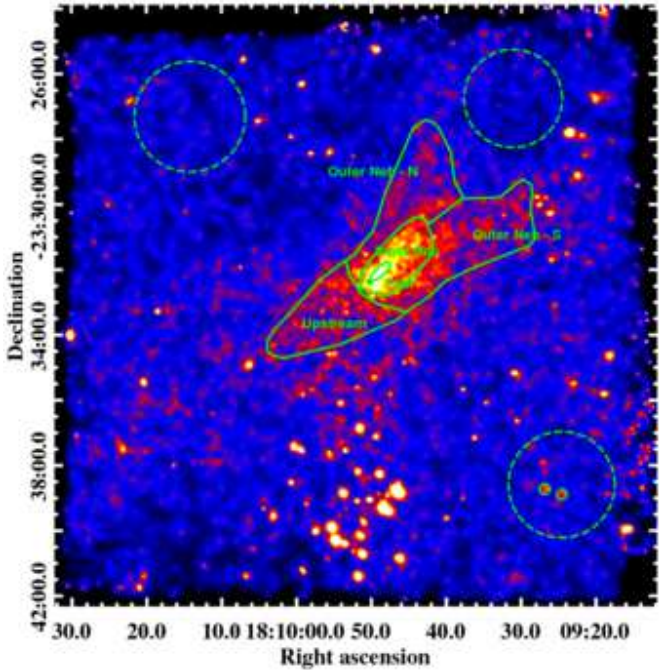}
\caption{
Merged 0.5-7 keV image showing the spectral extraction regions and background
regions (dotted).  The torus region is too small to be clearly seen, but 
is a $3.5\arcsec \times 5\arcsec$ ellipse centered on the pulsar
and excising the inner $2\arcsec$. 
}
\end{figure}

For the point source (neutron star), we first attempted a joint BB+PL
fit, but found that the PL index was driven to very soft values $\Gamma>6$,
indicating a poor fit to the thermal spectrum. If fixed at a more physical
$\Gamma=2$ the fit is unacceptable. However, if we adopt a neutron star 
atmosphere model (Sherpa model xsnsa) we obtain a reasonable reduced $\chi^2$
and powerlaw index.  This model includes a number of 
parameters, three of which we fix: neutron star gravitation mass = $1.5 M_{\odot}$,
neutron star radius = 10 km, neutron star magnetic field = $10^{13}$ G. The 
remaining variables are the effective temperature and model normalization.
The model normalization provides an estimate of the pulsar distance, which
for the fitted parameters yields $d=0.3_{-0.1}^{+1.0}$ kpc. 
This distance estimate is not very meaningful since, as usual,
the thermal emission likely has contributions from a heated polar
cap or soft magnetospheric power law. With the addition of such components the inferred
distance increases; it is in any event consistent with the $N_H$
constraint and the \citet{oka99} distance estimate.

	The diffuse spectral indices are not atypical of hard PWN emission.
We do not see any clear systematic softening as one moves farther from the pulsar.
However both the `Upstream' region and the `Outer Neb -S' appear significantly
softer than the rest of the nebula. This is especially true if $N_H$ is allowed
to vary, when somewhat larger $\Gamma$ values are preferred. The upstream emission 
is difficult to explain given the detected proper motion of the pulsar, unless it 
represents a foreground or background structure.  This leads to the
tentative suggestion that the hard-spectrum PWN is superposed on a softer
background arc composed of the `Upstream' and `Outer Neb -S' regions.
This emission may be unrelated to the PWN. A plausible origin is a reverse 
shock structure in the SNR.
Unfortunately we lack adequate S/N to confirm the spectral differences or even to
test whether a thermal fit is more suitable than a power-law model. 
In contrast, the `Outer Neb -N' region is formally
very hard, quite distinct to the Southern branch. Again the limited
counts preclude any detailed spectral study.

\begin{table*}[htb!]
\caption{X-Ray Spectral Fits}
\label{tab:spec}
\centering
\begin{tabularx}{1.0\textwidth}{lZZZZZZZ}
\hline \hline
Region \T \B & $N_{H}^a$ & $\Gamma$ / Log $T_\mathrm{eff}$ & Abs. Flux$^b$ &  Unabs. Flux$^b$  & $\chi^2$/d.o.f. \\
\hline
Trail+Inner\T \B& $1.31_{-0.18}^{+0.20}$ & $1.83 \pm 0.16$ & $6.3_{-1.2}^{+1.6}$ & $11.9_{-2.3}^{+3.1}$ & 112.9/199 \\
\hline
PSR\T \B & $1.3^c$ & $2.00\pm0.54$ / $5.9 \pm 0.1$ & $0.8_{-0.4}^{+0.8}$ & $1.2_{-0.6}^{+1.1}$ & 12.7/26 \\
\hline
Torus \T & $1.3^c$ & $1.77_{-0.38}^{+0.41}$ & $0.3_{-0.1}^{+0.2}$ & $0.6_{-0.2}^{+0.3}$ & 5.3/8 \\
Trail & $1.3^c$ & $1.88_{-0.18}^{+0.19}$ & $0.9 \pm 0.2$ & $1.8_{-0.3}^{+0.4}$ & 14.9/29 \\
Inner Neb& $1.3^c$ & $1.79 \pm 0.09$ & $4.6_{-0.4}^{+0.5}$ & $8.4_{-0.8}^{+0.9}$ & 76.3/159 \\
Upstream & $1.3^c$ & $2.21_{-0.18}^{+0.19}$ & $2.3 \pm 0.4$ & $5.4_{-0.9}^{+1.0}$ & 93.0/152 \\
Upstream & $1.8_{-0.4}^{+0.5}$ & $2.65_{-0.41}^{+0.47}$ & $2.2_{-0.9}^{+1.8}$ & $12.9_{-5.4}^{+10.4}$ & 91.6/151 \\
Outer Neb& $1.3^c$ & $1.56 \pm 0.10$ & $8.4_{-0.9}^{+1.0}$ & $13.7_{-1.5}^{+1.6}$ & 196.6/348 \\
Outer Neb - S & $1.3^c$ & $2.19 \pm 0.14$ & $3.9 \pm 0.5$ & $9.0_{-1.1}^{+1.2}$ & 108.2/209 \\
Outer Neb - S & $1.8^c$ & $2.61 \pm 0.16$ & $3.7_{-0.5}^{+0.6}$ & $13.3_{-1.8}^{+2.0}$ & 113.0/209 \\
Outer Neb - N & $1.3^c$ & $1.00 \pm 0.17$ & $4.3_{-0.9}^{+1.0}$ & $5.8_{-1.2}^{+1.3}$ & 93.5/178 \\

\hline
\end{tabularx}
$^a${interstellar absorption $\times10^{22} \, \rm{cm}^{-2}$}

$^b${$0.5-7$\,keV fluxes in units of $10^{-13} \, {\rm erg\,cm^{-2}\,s^{-1}}$}

$^c${held fixed} 
\end{table*}

\section{Discussion and Conclusions}

	The displacement of $0.60\pm 0.13$ pixels over 10.95 years gives 
a proper motion of $\mu = 27 \pm 5\, {\rm mas \, yr^{-1}}$. This corresponds to a space
velocity of $231\pm 46 (d/1.8\,{\rm kpc})\, {\rm km\, s^{-1}}$, a modest young pulsar
space velocity.  Interestingly, the proper motion vector points back to the
birthsite of $l=7.53^\circ$, $b=-1.68^\circ$ inferred within the $\sim 1.5^\circ$
diameter radio shell G7.5$-$1.7 by \citet{roberts08}. This is $1300\arcsec$ from the
pulsar so, taken at face value, our proper motion implies a kinematic age of
$48 \pm 10$\,kyr. Given the characteristic age $\tau_c=68$\,kyr, one gets 
an initial spin period $P_0 \sim 104 \pm 20$\,ms, for a braking index $n=3$.

	As noted by \citet{roberts08}, the pulsar projected offset is only 
$\sim 45$\% of the SNR radius, so if the PWN lies in the SNR interior it
may be just passing from being `crushed' by the reverse shock to forming 
a well-defined bow shock \citep{vDK04}. Accordingly, we should not be too 
surprised that the morphology at the PWN apex (Figure \ref{fig:f1b}) is 
unclear. Certainly the bulk of the hard spectrum emission trails the 
pulsar, following the proper motion axis and the direction to the explosion site.
However, there are significant counts bracketing the point source and
extending $\sim 5\arcsec$ {\it transverse} to the pulsar motion. Since the
pulsar spin axis appears to correlate with the proper motion \citep{jet05, nr07} 
such transverse extension tends to be equatorial. It is tempting to infer
that the blocky PWN head is the result of an anisotropic pulsar wind,
concentrated in an equatorial torus, with a spin axis/Earth line-of-sight
angle $\zeta \ge 75^\circ$, i.e. a torus viewed at large inclination angle. 
It is interesting to compare with predictions for the observed $\gamma$-ray 
pulse, which is a fairly narrow double with peak separation $\Delta = 0.35$.
Examining the `Atlas' of \citet{rw10}, we see that $\gamma$-ray pulsars 
with $\Delta \approx 0.35$ should have  $70^\circ > \zeta > 80^\circ$ and 
magnetic axis inclination $\alpha < 60^\circ$ for outer-magnetosphere
dominated emission.  There is little phase space for such pulsars 
to be radio detected. TPC-type
models produce such $\Delta$ for a wide range of $\zeta < 70^\circ$, but
many of the solutions should be radio detected. Thus both model classes
are allowed, although an outer-gap type interpretation seems preferred.
A good measurement of $\zeta$ from a detailed map of the PWN head could check this
interpretation, but would require a rather long ACIS exposure.

	Looking ahead of this transverse structure, we see that the hard spectrum 
diffuse emission has a fairly abrupt cutoff at $\sim 1\arcsec$ (Figure 1b).
We can compare this with the standoff angle for an isotropic ram-pressure 
confined pulsar wind
\begin{eqnarray}\label{eqn:bs}
\theta_{BS} \approx \left( \frac{\dot E}{4\pi c\rho v^2 d^2} \right)^{1/2} 
        \approx 1.3^{\prime\prime} n^{-1/2} \, \, d_{1.8}^{-2},
\end{eqnarray}
for our measured proper motion, an ambient number density $n\,{\rm cm^{-3}}$ and  
a distance 1.8\,kpc. It seems that a small standoff is not unexpected, but
given the apparent anisotropy in the wind momentum and likely anisotropy
in the SNR interior, no strong conclusions should be drawn.

	Of course there is emission `Upstream', ahead of the pulsar motion. The chip
gap spanned this region in the 2000 data so it was difficult to draw morphological
conclusions. In the combined data it seems that this emission is morphologically
distinct to, and substantially softer than, the PWN trail. This larger-scale soft emission
is also seen in archival {\it XMM} and {\it ASCA} data where it also appears
distinct from the harder PWN trail. In our data it appears to connect to the 
"Outer Neb - S" region, so we can posit that this extended softer emission
may be a background or foreground structure associated with the host SNR. 
If the absorption is left free, these regions
seem to prefer a slightly higher $N_H$ than the central PWN. Given the extensive
patchy molecular gas and obscuration in this region, these small differences
are not particularly telling.

\bigskip
\bigskip

We have examined a new \emph{Chandra} ACIS exposure of PSR J1809$-$2332
to study the fine scale X-ray morphology. Comparison with the 2000 exposure
yields a $>4\sigma$ detection of the pulsar proper motion and supports
the association with SNR G7.5$-$1.7.  The PWN has slightly extended emission at the
apex, somewhat larger than expected from a bow shock. While this suggests
that the pulsar wind is concentrated transverse to its velocity, the
possibility of anisotropies in the SNR interior discourages strong conclusions
and such disturbed morphology is not unexpected. Overall, however, these
new data support a basic picture for the PSR J1809$-$2332 system:
a $\sim 50$ky pulsar viewed near the spin equator, but with small magnetic
inclination so that the radio beam misses the Earth. The pulsar, traveling
at $\sim 230 {\rm km \, s^{-1}}$, is followed by a trail of hard PWN
X-ray emission and is approaching the outer region of its composite
host SNR located at $d \le 2$\,kpc. 

\acknowledgements
This work was supported in part by \emph{Chandra}
grant G01-12073X issued by the Chandra X-Ray Center, which is
operated by the Smithsonian Astrophysical Observatory for and
on behalf of the National Aeronautics Space Administration 
under contract NAS8-03060.

\end{document}